\documentclass[12pt]{article}
\usepackage{graphicx,epsfig,color}
\usepackage{cite}
\usepackage[english]{babel}
\usepackage{amssymb,amsfonts,amsmath}
\usepackage{verbatim}
\usepackage{tikz}
\usepackage{bbold}
\usepackage{hyperref}
\usepackage{float}

\newcommand{\cA}{{\cal A}}  
  \newcommand{\cD}{{\cal D}}

\newcommand{\cM}{{\cal M}}  \newcommand{\cN}{{\cal N}}
\newcommand{\cO}{{\cal O}}  
  
  \newcommand{\cT}{{\cal T}}

  \newcommand{\cZ}{{\cal Z}}

\newcommand{\be}{\begin{equation}} \newcommand{\ee}{\end{equation}}
\newcommand{\bea}{\begin{eqnarray}} \newcommand{\eea}{\end{eqnarray}}
\newcommand{\beann}{\begin{eqnarray*}}  \newcommand{\eeann}{\end{eqnarray*}}
\newcommand{\bfig}{\begin{figure}} \newcommand{\efig}{\end{figure}}
\newcommand{\ba}{\begin{array}} \newcommand{\ea}{\end{array}}
\newcommand{\bcen}{\begin{center}} \newcommand{\ecen}{\end{center}}
\newcommand{\btab}{\begin{tabular}} \newcommand{\etab}{\end{tabular}}

\def\tr{\operatorname{tr\:}}

\newcommand{\bra}[1]{\langle #1|}
\newcommand{\ket}[1]{|#1\rangle}
\newcommand{\vev}[1]{\left\langle{#1}\right\rangle}

\newtheorem{Proposition}{Proposition}[section]

\newtheorem{Theorem}{Theorem}[section]
\newtheorem{Lemma}{Lemma}[section]
\newtheorem{Corrolary}{Corrolary}[section]

\newcommand{\bp}{\begin{Proposition}}   \newcommand{\ep}{\end{Proposition}}
\newcommand{\bt}{\begin{Theorem}}   \newcommand{\et}{\end{Theorem}}
\newcommand{\bl}{\begin{Lemma}}     \newcommand{\el}{\end{Lemma}}
\newcommand{\bc}{\begin{Corrolary}} \newcommand{\ec}{\end{Corrolary}}

\def\ep{\epsilon}

\usepackage[top=2cm, bottom=2cm, left=2cm, right=2cm]{geometry}	

\numberwithin{equation}{section}

\begin{document}

\begin{flushright}
HIP-2019-35/TH
\end{flushright}

\begin{center}

\centering{\Large {\bf Scattering length from holographic duality}}

\vspace{8mm}

\renewcommand\thefootnote{\mbox{$\fnsymbol{footnote}$}}
Carlos Hoyos,${}^{1,2}$\footnote{hoyoscarlos@uniovi.es}
Niko Jokela,${}^{3,4}$\footnote{niko.jokela@helsinki.fi} and
Daniel Logares${}^{1,2}$\footnote{dani.logares@gmail.com} 

\vspace{4mm}
${}^1${\small \sl Department of Physics} \\
{\small \sl Universidad de Oviedo} \\
{\small \sl c/ Federico Garc\'{\i}a Lorca 18, ES-33007 Oviedo, Spain} 

\vspace{2mm}
\vskip 0.2cm
${}^2${\small \sl Instituto Universitario de Ciencias y Tecnolog\'{\i}as Espaciales de Asturias (ICTEA)}\\
{\small \sl  Calle de la Independencia, 13, 33004 Oviedo, Spain}

\vspace{2mm}
\vskip 0.2cm
${}^3${\small \sl Department of Physics} and ${}^4${\small \sl Helsinki Institute of Physics} \\
{\small \sl P.O.Box 64} \\
{\small \sl FIN-00014 University of Helsinki, Finland} 

\end{center}

\vspace{8mm}

\renewcommand\thefootnote{\mbox{\arabic{footnote}}}

\begin{abstract}
\noindent  Interesting theories with short range interactions include QCD in the hadronic phase and cold atom systems. The scattering length in two-to-two elastic scattering process captures the most elementary features of the interactions, such as whether they are attractive or repulsive. However, even this basic quantity is notoriously difficult to compute from first principles in strongly coupled theories. We present a method to compute the two-to-two amplitudes and the scattering length using the holographic duality. Our method is based on the identification of the residues of Green's functions in the gravity dual with the amplitudes in the field theory. To illustrate the method we compute a contribution to the scattering length in a hard wall model with a quartic potential and find a constraint on the scaling dimension of a scalar operator $\Delta > d/4$. For $d< 4$ this is more stringent than the unitarity constraint and may be applicable to an extended family of large-$N$ theories with a discrete spectrum of massive states. We also argue that for scalar potentials with polynomial terms of order $K$, a constraint more restrictive than the unitarity bound will appear for $d<2K/(K-2)$.
\end{abstract}



\newpage

\section{Introduction}

Strong nuclear forces are well described at high energies by the perturbative expansion of quantum chromodynamics (QCD), the microscopic theory that has as dynamical degrees of freedom quarks and gluons. At low energies QCD becomes strongly coupled and an effective description becomes a necessity. Mesons and baryons (hadrons) are the observed physical bound states at low energies and a standard approach is to construct an effective field theory with hadrons as dynamical degrees of freedom; for reviews on this approach, see \cite{Kaplan:2005es,Epelbaum:2008ga,Hammer:2016xye}. Some properties of the underlying QCD theory, especially symmetries, are valuable guides in constructing the effective action for hadrons, and numerical lattice computations offer a first principles methodology to compute the spectrum. Due to confinement, strong interactions are short-ranged and in many cases can be approximated by contact interactions, which are determined by the scattering length measured in the experiment, or alternatively extracted from finite size corrections in lattice QCD \cite{Luscher:1986pf,Luscher:1990ux,Rummukainen:1995vs}. In general, there are also interactions within the hadronic theory produced by hadron exchange. As experiments are necessarily made in some fixed conditions, the extrapolation of the effective action to other regimes, such as large baryon densities, breaks down at some point.

Hadronic matter is not the only example of a physical system with short-range interactions. Those abound in condensed matter systems, for instance in the descriptions of the effective potential for cold atoms at unitarity \cite{Eagles:1969zz,Nozieres:1985zz,Leggetbook}. In many cases, the details of the interactions for a given channel are not very important, and for an effective description at low energies it is enough to know the scattering length $a_s$ and the range of interaction. Generically, a negative scattering length $a_s<0$ implies that the interaction is attractive, while a positive value $a_s>0$ may signal either a repulsive interaction (for small $a_s$) or the existence of bound states in an attractive potential (for large $a_s$). 

We argue that a similar ``hadronic'' effective description can be obtained in certain holographic models, those with a mass gap and a discrete spectrum of excitations for gauge-invariant operators. Examples with a known field theory dual consist of the Witten Yang-Mills model \cite{Witten:1998zw} and its various generalizations: the Sakai-Sugimoto model \cite{Sakai:2004cn,Sakai:2005yt}, the Klebanov-Strassler model \cite{Klebanov:2000hb}, and the $\cN=1^*$ SYM dual to the  GPPZ geometry \cite{Girardello:1999bd}. In this paper, we will restrict to states created by operators that have a dual description as classical fields in gravity, typically glueballs and mesons. Some states, baryons for instance, map to heavy objects such as wrapped branes on the gravity side \cite{Witten:1998xy}. The effective potential for these types of objects has been studied for instance in the Sakai-Sugimoto model using solitonic configuration on flavor branes \cite{Hashimoto:2008zw,Hashimoto:2009ys}. An analysis similar to ours is the calculation of glueball decay rates using three-particle scattering amplitudes in the GPPZ model \cite{Mueck:2004qg}, while for very high energy scattering a description in terms of strings may be more appropriate \cite{Polchinski:2001tt}.

A reason why we find this kind of analysis useful, in particular in the context of applications of the duality to QCD or condensed matter systems, is that it can help to understand the effect of parameters in the gravitational action on the properties on the field theory side. The scattering length is one such physical observable and, as we will show, will crucially depend on the coupling constants in the bulk scalar potential. One can therefore learn how varying the parameter values in the scalar potential determine the strength of the effective interactions and whether they are attractive or repulsive. This makes the gravitational action less of a `black box' and helps in determining what kind of an action mimics the desired physics, given the field content. In addition, the scattering length calculations will be useful to further constrain the specific gravitational actions of models aspired to give reliable descriptions of QCD. 

Furthermore, the gravitational action describes both the hadronic phase with confinement and a deconfined phase, where quarks and gluons should be the dynamical degrees of freedom, although they cannot be directly observed as they are not gauge-invariant fields. On the other hand, physical quantities, like the equation of state (see, {\emph{e.g.}}, \cite{Hoyos:2016cob,Ecker:2017fyh}), in the deconfined phase depend on the same parameters of the gravitational action as the scattering lengths. Therefore, our analysis establishes a direct link between hadronic physics and the physics of quarks and gluons at low energies, something that seems out of reach using ordinary field theory methods.

The outline of the paper is the following. We will start in Sec.~\ref{sec:scat} reviewing the derivation of the scattering length from the four-point function in four spacetime dimensions. In Sec.~\ref{sec:holo} we will present the method to compute the scattering amplitude and the scattering length between four scalar particles using holography. We will allow a general value of the spacetime dimension and show that it is enough to extract the residues of poles in Green's functions of classical fields in the gravitational theory. Finally, for illustration 
purposes, in Sec.~\ref{sec:hw} we will apply the method to the hard wall model, that has been proposed as a toy model of confinement in QCD \cite{Erlich:2005qh} and gapped states in condensed matter systems \cite{Sachdev:2011ze}. We discuss our results and possible future directions in Sec.~\ref{sec:discuss}.

\section{Scattering length}\label{sec:scat}

We will be studying large-$N$, strongly coupled theories in four spacetime dimensions with a gapped, discrete spectrum and in particular focus on a scalar gauge-invariant single-trace operator $\cO$. Examples include a glueball operator such as $\cO=\tr F^2$, where $F$ the field strength of the gauge fields, or $\cO=\overline{q} q$, where $q$ corresponds to a quark operator. When the operator acts on the vacuum state, it creates a scalar excitation belonging to a tower of states all with the same quantum numbers and different masses $m_a$, $a=1,2,3,\ldots$. In the large-$N$ limit these states behave approximately as almost free particles, with interactions that can be treated perturbatively in a $1/N$ expansion.

The tower of massive states is captured by the two-point function of the scalar operator, in momentum space $p=(\omega,\mathbf{k})$,
\be
\vev{\cO(p)\cO(p')}= (2\pi)^4\delta^{(4)}(p+p')G(p) \ .
\ee
When the momentum is put on-shell at one of the masses $\omega\to \varepsilon_a(\mathbf{k})=\sqrt{\mathbf{k}^2+m_a^2}$, there is a pole singularity in the two-point function 
\be
G(p)\sim  \frac{-i Z_a}{p^2+m_a^2}+\ldots \ .
\ee
We can extract the residues of these poles from a LSZ-like reduction formula
\be
Z_a=i\lim_{\omega\to \varepsilon_a} (p^2+m_a^2)G(p) \ .
\ee
Since the width of these states is suppressed in the large-$N$ limit, it is possible to define asymptotic non-interacting states and compute the scattering matrix. The S-matrix can be split in the non-interacting part and the interacting part, or $T$-matrix,
\be
S=\mathbb{1}+i\cT \ .
\ee
In a two-to-two scattering of the spin-zero particles with masses $m_n$ and spatial momenta $\mathbf{k}_n$, $n=1,2,3,4$, the $T$-matrix element
\be
i\cT_{12\to 34}=\bra{ m_3, \mathbf{k}_3; \, m_4, \mathbf{k}_4}\, i\cT\, \ket{ m_1,\mathbf{k}_1;\, m_2,\mathbf{k}_2}
\ee
can be obtained from a connected four-point function using the LSZ reduction formula ($\varepsilon_n\equiv \sqrt{\mathbf{k}_n^2+m_n^2}$)
\be\label{eq:Tmatrix}
i\cT_{12\to 34}\equiv \lim_{\omega_n \to \varepsilon_n} \left(\prod_{n=1}^4 \frac{p_n^2+m_n^2}{Z_n^{1/2}}\right)\vev{\cO(p_1)\cO(p_2)\cO(p_3)\cO(p_4)}_c \ .
\ee

To define the scattering length one should first recall that momentum states are defined with different normalization in the relativistic and non-relativistic approaches. 
Taking this into account, we extract the following factor from the matrix elements 
\be
S_{12\to 34}=4(\varepsilon_1\varepsilon_2\varepsilon_3\varepsilon_4)^{1/2}\widetilde{S}_{12\to 34} \ .
\ee
Here $\widetilde{S}_{12\to 34}$ will become the non-relativistic scattering matrix. Let us now focus on the case of elastic scattering between two particles of possibly different masses $m_1=m_3$ and $m_2=m_4$. Momentum conservation implies that the $T$-matrix contribution should take the form
\be
i\widetilde{\cT}_{12\to 34}=(2\pi)^4\delta(\varepsilon_1+\varepsilon_2-\varepsilon_3-\varepsilon_4 )\delta^{(3)}(\mathbf{k}_1+\mathbf{k}_2-\mathbf{k}_3-\mathbf{k}_4) i\widetilde{\cM} \ .
\ee
In the non-relativistic limit $\varepsilon_n\approx m_n+\frac{\mathbf{k}_n^2}{2 m_n}$. Going to the center of mass frame $\mathbf{k}_1=-\mathbf{k}_2=\mathbf{k}$ and integrating over the final momenta of the particle of mass $m_1$ or $m_2$ 
\be
 i\widetilde{\cT}_{\mathbf{k}\to \mathbf{k}'} = i (2\pi) \delta\left(\frac{{\mathbf{k}'}^2}{2 m_{12}}-\frac{{\mathbf{k}}^2}{2 m_{12}} \right) \widetilde{\cM}(\mathbf{k}'-\mathbf{k}) \ ,
\ee
where $\mathbf{k}'=\mathbf{k}_3=-\mathbf{k}_4$, $m_{12}=m_1 m_2/(m_1+m_2) $ is the reduced mass and we have assumed that the scattering amplitude depends only on the exchanged momentum between the particles. The formula above can be understood as the Born approximation to the scattering of a particle of mass $m_{12}$ in a potential, the Fourier transform of the potential being $\widetilde{V}(\mathbf{q})=- \widetilde{\cM}(\mathbf{q)}$. It follows that the scattering amplitude of the outgoing wave is
\be
 f(\mathbf{q})=\frac{m_{12}}{2\pi} \widetilde{\cM}(\mathbf{q}) \ .
\ee
Therefore, the scattering length is
\be\label{eq:as}
 a_s=-\lim_{\mathbf{q}\to \mathbf{0}}\frac{m_{12}}{2\pi} \widetilde{\cM}(\mathbf{q}) \ .
\ee

\section{Scattering amplitude in holography}\label{sec:holo}

We will assume that a large-$N$ strongly coupled gauge theory in $d$ spacetime dimensions has a holographic dual description.  The dual will in general consist of Einstein gravity in $d+1$ dimensions coupled to a scalar field $\Phi$ dual to $\cO$
\be\label{eq:haction}
 S = \frac{1}{16\pi G_N}\int d^{d+1} x \sqrt{-g}\left(R-(\partial\Phi)^2-2V(\Phi)\right) \ .
\ee
In order for the theory to be well-defined in the UV, we will assume that the potential has a critical point $V'(0)=0$. In this case there is a pure $AdS_{d+1}$ solution, dual to a UV fixed point, where the scalar is constant $\Phi=0$. The asymptotic metric is
\be\label{eq:adsmetric}
 \overline{g}_{MN}=\frac{L^2}{z^2}\eta_{MN} \ , \ z\to 0 \ ,
\ee
where $\eta_{MN}$ is the flat $d$-dimensional Minkowski metric. We will use capital Latin letters for spacetime indices in the bulk of the gravity dual and Greek letters for the field theory directions.

The potential expanded around the UV critical point will have the form
\be\label{eq:scalarpot}
 V(\Phi)\simeq -\frac{d(d-1)}{2L^2}+\frac{1}{2}m^2\Phi^2+\frac{v_4}{4L^2}\Phi^4+\ldots \ .
\ee
The mass $m$ of the scalar determines the scaling dimension $\Delta$ of the dual operator $m^2L^2=\Delta(\Delta-d)$.  Let us introduce the following parametrization of the scaling dimension of the dual operator
\be
 \Delta=\frac{d}{2}+\nu \ .
\ee
Then, the mass is $m^2 L^2=\nu^2-\frac{d^2}{4}$. The quartic term introduces a contact interaction for the scalar in the bulk, proportional to $v_4$. 

With this action, the asymptotic expansion of the scalar close to the $AdS_{d+1}$ boundary is in general of the form
\be\label{eq:scalarexp}
 \Phi(x,z)=\left(\frac{z}{L}\right)^{\frac{d}{2}-\nu}\left[A_0(x)+\frac{z}{L} A_1(x)+\ldots+\left(\frac{z}{L}\right)^{2\nu} \left(C_\nu(x)+\frac{z}{L} C_{\nu,1}(x)+\ldots\right)\right] \ ,
\ee
where the first part corresponds to the non-normalizable solution and the second part to the normalizable solution. If $\frac{d}{2}+\nu$ is an integer, then the expansion of the non-normalizable solution may include also logarithmic terms $\log(z/L)$. 

The expansion of the metric is
\be\label{eq:metricexp}
g_{MN}(x,z)=\frac{L^2}{z^2}\left( \eta_{MN}+\frac{z^2}{L^2} H_{2\, MN}(x)+\ldots+\frac{z^d}{L^d} H_{d\,MN}(x)+\ldots\right) \ .
\ee
The equations of motion fix the coefficients in the non-normalizable solution of the scalar $A_1$ etc., and the coefficients of the metric $H_{n\, MN}$ for $n<d$ (as well as the coefficients of logarithmic terms). All these coefficients are local functionals of $A_0$ and its derivatives along the field theory directions. 

\subsection{$n$-point functions}

In order to compute the $T$-matrix element \eqref{eq:Tmatrix} we need both the residues of the poles in the two-point and the four-point functions, in the limit where the external momenta are on-shell. The correlators can be computed from the one-point function in the presence of an external source $j$ for the scalar operator 
\be
 \vev{\cO(x)}_j=\frac{\int \cD\cA\, \cO(x) \,e^{i S_{\rm YM}+i\int d^dx j(x) \cO(x)}}{\int \cD\cA\,e^{i S_{\rm YM}+i\int d^dx j(x) \cO(x)}} \ .
\ee
Here $S_{\rm YM}$ is the action of the dual field theory, depending on the fields $\cA$ and $\cD\cA$ is the field theory path integral measure. Note that the operator $\cO(x)$ is a functional of the fields $\cA$. Higher order correlation functions are obtained from the one-point function as
\be\label{eq:npointf}
 \vev{\cO(x_1)\cdots \cO(x_n)}=\frac{1}{n}\sum_{k=1}^n \prod_{\overset{ a=1}{ a\neq k}}^{n} \left(\frac{1}{i}\frac{\delta}{\delta j(x_a)}\right) \vev{\cO(x_k)}_j \Big|_{j=0} \ .
\ee

An external source is realized in the gravity dual as a boundary condition for the scalar field. This fixes $A_0(x)=L^{\frac{d}{2}-\nu} j(x)$:
\be\label{eq:bcphi}
 \Phi(x,z\to 0)= z^{\frac{d}{2}-\nu} j(x) \ ,
\ee
The one-point function can be computed using the canonical momentum of the dual scalar $\Phi$ \cite{Papadimitriou:2004ap,Papadimitriou:2004rz}
\be
 \pi_\Phi=\sqrt{-g} g^{zz} \partial_z\Phi \ .
\ee
The regularized momentum $\pi_\Phi^{reg}=\pi_\Phi+\pi_\Phi^{c.t.}$ can be obtained from a variation of the action including counterterms. The one-point function is obtained as the finite result from \cite{Bianchi:2001de,Papadimitriou:2004rz} 
\be\label{eq:renvev}
 \vev{\cO(x)}_j=\cN\, \frac{1}{L^{\frac{d}{2}+\nu-1}}\lim_{z\to 0}\left(\frac{z}{L}\right)^{\frac{d}{2}-\nu} \pi_\Phi^{reg}=\cN \frac{2\nu C_\nu(x)}{L^{\frac{d}{2}+\nu}}+\text{local terms} \ .
\ee
In the special case $\nu=0$, the overall coefficient is $2$ instead of $2\nu$. The dimensionless factor $\cN=L^{d-1}/(16 \pi G_N)$ is proportional to the number of degrees of freedom of the field theory, $\cN\sim N^2$ for a large-$N$ gauge theory in $d=4$. The local terms are proportional to powers of $j$ or its derivatives and are in general scheme-dependent (terms depending on $\log j$ might be present as well). They will drop from the LSZ reduction formula, so we can neglect them.

In order to compute the correlators we expand around a background solution and solve the equations of motion derived from the action \eqref{eq:haction} with the boundary condition \eqref{eq:bcphi}. We then evaluate the solution on the regularized momentum and find the expectation value from \eqref{eq:renvev}. We construct the classical solution by introducing a small parameter $\epsilon$ and rescaling the source  $j\to \epsilon j$. Then we can do an expansion in powers of $\epsilon$ 
\be\label{eq:epsexp}
 \Phi=\phi+\epsilon \phi^{(1)}+\epsilon^2\phi^{(2)}+\ldots \ , \ g_{MN}=h_{MN} +\epsilon h^{(1)}_{MN}+\epsilon^2 h_{MN}^{(2)}+\ldots \ .
\ee
In this expansion $\phi$, $h_{MN}$ are the background fields, $\phi^{(1)}$ satisfy the boundary condition \eqref{eq:bcphi}, and the remaining terms should not change the boundary conditions for the scalar field of the metric. These requirements imply
\be
 \lim_{z\to 0} \frac{1}{z^{\frac{d}{2}-\nu}} \phi^{(n)}=0, \ n> 1,\ \ \ \lim_{z\to 0} z^2 h^{(m)}_{MN}=0\ , \ m>0 \ .
\ee
Each term in the expansion is a homogeneous functional of the source $j$, and we will assume that the background only depends on the radial coordinate $\partial_\mu \phi=\partial_\mu h_{MN}=0$. For the computation of the two- and four-point functions we just need to expand \eqref{eq:renvev} up to $O(\epsilon^3)$. 

At each order, the solution will be obtained as the convolution of a Green's function with the sources
\bea
 \phi^{(n)}(z,x) & = & \int d^d x_1\cdots d^d x_n\; G^{(n)}(z,x;x_1, \ldots, x_n) j(x_1)\cdots j(x_n) \\
 h_{MN}^{(n)}(z,x) & = & \frac{L^2}{z^2}\int d^d x_1\cdots d^d x_n G_{MN}^{(n)}(z,x;x_1, \ldots, x_n) j(x_1)\cdots j(x_n) \ .
\eea
The boundary conditions demand that
\bea
 \lim_{z\to 0} G^{(1)}(z,x; x_1) & \sim & z^{\frac{d}{2}-\nu}\delta^{(d)}(x-x_1) \\ 
 \lim_{z\to 0} \frac{1}{z^{\frac{d}{2}-\nu}}G^{(n)}(z,x; x_1, \ldots, x_n) & = & 0 \\ 
 \lim_{z\to 0} G_{MN}^{(n)}(z,x; x_1, \ldots, x_n)  & = & 0 \ .
\eea
It is convenient to perform the Fourier transform and work in the momentum space, 
\bea
 \widetilde{\phi}^{(n)}(z,p) & = & \int \frac{d^d p_1}{(2\pi)^d}\cdots \frac{d^d p_n}{(2\pi)^d}\; \widetilde{G}^{(n)}(z,p;-p_1, \ldots, -p_n) \widetilde{\text{\em \j}}(p_1)\cdots \widetilde{\text{\em \j}}(p_n) \\
 \widetilde{h}_{MN}^{(n)}(z,p) & = & \frac{L^2}{z^2}\int \frac{d^d p_1}{(2\pi)^d}\cdots \frac{d^d p_n}{(2\pi)^d} \widetilde{G}_{MN}^{(n)}(z,p;-p_1, \ldots,- p_n) \widetilde{\text{\em \j}}(p_1)\cdots \widetilde{\text{\em \j}}(p_n) \ .
\eea
Expanding for $z\to 0$, we find
\bea
 \widetilde{G}^{(1)}(z,p; -p_1) & \simeq & z^{\frac{d}{2}-\nu} (2\pi)^d \delta^{(d)}(p-p_1)+\ldots+\left(\frac{z}{L}\right)^{\frac{d}{2}+\nu} \widetilde{G}^{(1)}_\nu(p;-p_1)+\ldots \\
 \widetilde{G}^{(n)}(z,p; -p_1,\ldots,-p_n) & \simeq & \left(\frac{z}{L}\right)^{\frac{d}{2}+\nu} \widetilde{G}^{(n)}_\nu(p;-p_1,\ldots,-p_n)+\ldots \ .
\eea
Therefore, the contribution to the coefficient that determines the one-point function in \eqref{eq:renvev} is
\be\label{eq:nc3}
 \widetilde{C}_\nu^{(n)}(p)=\int \frac{d^d p_1}{(2\pi)^d}\cdots \frac{d^d p_n}{(2\pi)^d}\; \widetilde{G}^{(n)}_\nu(p;-p_1, \ldots, -p_n)  \widetilde{\text{\em \j}}(p_1)\cdots \widetilde{\text{\em \j}}(p_n) \ .
\ee

\subsection{Residues and scattering amplitude}

From \eqref{eq:npointf} and \eqref{eq:nc3} we can deduce the following expressions for the two- and four-point functions
\bea
\vev{\cO(p_1) \cO(p_2)} & = & -\frac{i \nu\cN}{L^\Delta}\left(\widetilde{G}^{(1)}_\nu(p_1;-p_2)+\widetilde{G}^{(1)}_\nu(p_2;-p_1)\right) \\
\vev{\cO(p_1) \cO(p_2)\cO(p_3)\cO(p_4)} & = & \frac{i\nu \cN}{2L^\Delta}\sum_{k=1}^4 \widetilde{G}_\nu^{(3)}(p_k ;- p_{n_1},-p_{n_2},-p_{n_3})\Big|_{n_i=1,2,3,4; n_i\neq k} \ .
\eea
In fact we will not need to compute the full Green's functions in the gravity side, it will be enough to keep track of the leading on-shell singularities. When the momenta of the external sources are taken close to the value of the masses of the scalar excitations we will find that the kernels are of the form
\bea
 \widetilde{G}^{(1)}(z;p_1;-p_2) & \simeq & (2\pi)^d \delta^{(d)}(p_1-p_2) \frac{\Gamma^{(1)}_a(z,p_1) }{p_1^2+m_a^2} \\
 \widetilde{G}^{(3)}(z,p_1;-p_2,-p_3,-p_4) & \simeq & (2\pi)^d \delta^{(d)}(p_1-p_2-p_3-p_4) \nonumber \\
  & & \times \frac{\Gamma^{(3)}_{a; bcd}(z,p_1,-p_2,-p_3,-p_4)}{(p_1^2+m_a^2)(p_2^2+m_b^2)(p_3^2+m_c^2)(p_4^2+m_d^2)} \ .
\eea
The residues can be directly read from the boundary expansion $z\to 0$ by isolating the poles
\be\label{eq:ZM}
 \Gamma^{(1)}_a\simeq z^{\frac{d}{2}+\nu} \frac{Z_a}{2\nu\cN}+\ldots,\ \ \Gamma^{(3)}_{a; bcd}\simeq z^\Delta \frac{\cZ_{a; bcd}}{2\nu\cN}+\ldots \ .
\ee
The (relativistic) scattering amplitude is obtained by symmetrizing the residue in the four-point kernel coefficient
\be\label{eq:Mabcd}
 \cM_{abcd}=\frac{1}{4}\sum_{k=a,b,c,d} \frac{\cZ_{k; n_1n_2n_3}}{(Z_a Z_b Z_c Z_d)^{1/2}}\Big|_{n_i=a,b,c,d; n_i\neq k} \ .
\ee
For $d=4$ spacetime dimensions we can compute the scattering length for elastic scattering between two particles of masses $m_a$ and $m_b$. Using the non-relativistic approximation and taking the zero momentum limit in the center of mass frame, 
\be\label{eq:ash}
 a_s=-\frac{\cM_{aabb}(\mathbf{k}=\mathbf{k}'=0)}{8\pi(m_a +m_b)} \ .
\ee

\section{Hard wall model}\label{sec:hw}

We will illustrate our approach with a simple example. We will treat the scalar field as a probe, neglecting the backreaction on the metric, \emph{i.e.}, it will remain fixed in this case. For simplicity, we consider $AdS_{d+1}$ \eqref{eq:adsmetric} metric in the whole bulk, but in order to mimic confinement we will introduce a hard wall \cite{Erlich:2005qh} at a finite value of the radial coordinate $z$. This means that the space abruptly ends, with the radial coordinate taking values on the interval $0<z<1/\Lambda$. A Dirichlet condition for the fields at the hard wall makes the spectrum of fluctuations discrete and gapped, with the gap set by the IR scale $\Lambda$.

The equations of motion for the scalar field are ($\square\equiv \eta^{\mu\nu}\partial_\mu\partial_\nu$)
\be
 z^2\partial_z^2\Phi-(d-1) z\partial_z\Phi+z^2 \square \Phi-L^2V'(\Phi)=0 \ .
\ee
We will take the scalar potential to be of the form \eqref{eq:scalarpot}, and do the expansion \eqref{eq:epsexp} for the scalar field with a vanishing background field $\phi=0$. The equation of motion at $O(\epsilon^n)$ is
\be
 z^2\partial_z^2\phi^{(n)}-(d-1) z\partial_z\phi^{(n)}+z^2 \square \phi^{(n)}-m^2L^2\phi^{(n)}=W^{(n)} \ ,
\ee
where the inhomogeneous terms read, up to $O(\epsilon^3)$,
\be
 W^{(1)}=W^{(2)}=0,\ \ W^{(3)}=v_4 (\phi^{(1)})^3 \ .
\ee
Fixing the mass of the scalar to $m^2L^2=\nu^2-\frac{d^2}{4}$, the boundary conditions at the hard wall and the asymptotic $AdS$ boundary are 
\be
 \phi^{(n)}\Big|_{z=1/\Lambda}=0,\ \ \phi^{(1)}\underset{z\to 0}{\sim} z^{\frac{d}{2}-\nu} j(x), \ \ \phi^{(n)}\underset{z\to 0}{\sim} O(z^{\frac{d}{2}+\nu}), \ n>1 \ .
\ee
With these conditions $\phi^{(2)}=0$. We will now go to momentum space via Fourier transformation
\be
 \widetilde{\phi}^{(1)}(p,z)=\int d^d x e^{-ip\cdot x} \phi^{(1)}(z,x) \ .
\ee
The equations of motion are
\be
 \left(z^2\partial_z^2-(d-1) z\partial_z+z^2 M^2+\frac{d^2}{4}-\nu^2\right)\widetilde{\phi}^{(n)}(p,z)=\widetilde{W}^{(n)}(p,z) \ ,
\ee
where we have defined $M^2=-p^2$ and 
\be
 \widetilde{W}^{(3)}(p,z)=(2\pi)^d \delta^{(d)}(p-p_1-p_2-p_3)v_4\int \frac{d^d p_1}{(2\pi)^d} \frac{d^d p_2}{(2\pi)^d} \frac{d^d p_3}{(2\pi)^d} \widetilde{\phi}^{(1)}(p_1,z)\widetilde{\phi}^{(1)}(p_2,z)\widetilde{\phi}^{(1)}(p_3,z) \ .
\ee
The solutions for $\widetilde{\phi}^{(1)}$ and $\widetilde{\phi}^{(3)}$ can be written as
\be
 \widetilde{\phi}^{(1)}(p,z)=K(p,z)\widetilde{\text{\em \j}}(p),\ \ \widetilde{\phi}^{(3)}(p,z)=\int_0^{1/\Lambda} dz_1\frac{L^{d+1}}{z_1^{d+1}}\Sigma(p,z,z_1) \widetilde{W}^{(3)}(p,z_1)\ . 
\ee
Where we have introduced the bulk-to-boundary ($K$) and the bulk-to-bulk ($\Sigma$) propagators. The equations of motion for each of the propagators is
\bea
 & & \left(z^2\partial_z^2-(d-1) z\partial_z+z^2 M^2+\frac{d^2}{4}-\nu^2\right)K(p,z)=0 \\
 & & \left(z^2\partial_z^2-(d-1) z\partial_z+z^2 M^2+\frac{d^2}{4}-\nu^2\right)\Sigma(p,z,z_1)=\frac{z^{d+1}}{L^{d+1}}\delta(z-z_1) \ .
\eea
They must satisfy the boundary conditions
\be
 K(p,1/\Lambda)=0\ , \ K(p,z) \underset{z\to 0}{\sim} z^{\frac{d}{2}-\nu}\ , \ \Sigma(p,1/\Lambda,z_1)=0\ ,\ \Sigma(p,z,z_1) \underset{z\to 0}{\sim} O(z^{\frac{d}{2}+\nu}) \ .
\ee
Then, the expansions for $z\to 0$ should be of the form
\bea
K(p,z)& \simeq & z^{\frac{d}{2}-\nu} +\cdots+\left(\frac{z}{L}\right)^{\frac{d}{2}+\nu}K_\nu(p)+\ldots \\ 
\Sigma(p,z,z_1) & \simeq & \left(\frac{z}{L}\right)^{\frac{d}{2}+\nu}\Sigma_\nu(p,z_1)+\ldots \ .
\eea
Direct comparison with \eqref{eq:nc3} shows that
\bea
 \tilde{G}^{(1)}_\nu(p;-p_1) & = & (2\pi)^d \delta^{(d)}(p-p_1)K_\nu(p_1) \nonumber\\
 \tilde{G}^{(3)}_\nu(p;-p_1,-p_2,-p_3) & = & (2\pi)^d \delta^{(d)}(p-p_1-p_2-p_3)\, v_4 \nonumber\\
  & & \times\int_0^{1/\Lambda} dz_1\frac{L^{d+1}}{z_1^{d+1}}\Sigma_\nu(p,z_1) K(p_1,z_1) K(p_2,z_1) K(p_3,z_1) \ .\label{eq:K3resHW}
\eea

\subsection{Bulk-to-boundary and bulk-to-bulk propagators}

The equation of motion that the propagators must satisfy can be put in Sturm-Liouville form. Multiplying the equation by $1/(\Lambda z)^{d+1}$, using as variable $u=\Lambda z$ and defining $\mu=M/\Lambda$,
\bea
 \partial_u \left( \frac{1}{u^{d-1}} \partial_u K(p,u)\right)+\left(\frac{\frac{d^2}{4}-\nu^2}{u^{d+1}}+ \frac{\mu^2}{u^{d-1}}\right) K(p,u) & = & 0 \\
 \partial_u\left( \frac{1}{u^{d-1}} \partial_u\Sigma(p,u,u_1)\right)+\left(\frac{\frac{d^2}{4}-\nu^2}{u^{d+1}}+ \frac{\mu^2}{u^{d-1}}\right) \Sigma(p,u,u_1) & = & \frac{\Lambda}{(L\Lambda)^{d+1}}\delta(u-u_1) \ .
\eea
The constant factor multiplying the delta function can be absorbed in the normalization of $\Sigma$. The eigenvalues of the associated Sturm-Liouville  problem are $\mu^2$, and the weight function is $w(u)=1/u^{d-1}$.

The bulk-to-boundary propagator is easy to find as a solution to the homogeneous equation satisfying the Dirichlet boundary condition and $K(p,u)\sim (u/\Lambda)^{\frac{d}{2}-\nu}$ when $u\to 0$:
\be
 K(p,u)=-\frac{\pi \mu^\nu}{2^\nu\Gamma(\nu)\Lambda^{\frac{d}{2}-\nu}}  u^{d/2}\left( Y_\nu(\mu u)-c_\mu J_\nu(\mu u)\right) \ , \ c_\mu=\frac{Y_\nu(\mu)}{J_\nu(\mu)}\ . 
\ee
Let us consider solutions to the homogeneous equation satisfying a Dirichlet condition at the hard wall and a normalizable condition at the $AdS$ boundary. These solutions only exist for a discrete set of values of $M$, corresponding to the zeroes of the Bessel function $J_\nu$:
\be
 \varphi_a(u)=c_a u^{d/2} J_\nu(\mu_a u) \ , \ c_a=\frac{\sqrt{2}}{J_{\nu+1}(\mu_n)}\ ,  \ J_\nu(\mu_a)=0 \ ,\ a=1,2,3,\ldots \ .
\ee
The values of $M_a=\mu_a \Lambda$ are the masses of glueball or meson states created by the operator dual to the scalar field. When we compute the residues of the two-point function and scattering amplitudes we will take the momenta to be on-shell on one of these masses. The solutions are normalizable as long as $\nu>-1$, which corresponds to the unitarity bound.

The functions $\varphi_a$ form an orthonormal set
\be
 \int_0^1 \frac{du}{u^{d-1}} \varphi_a(u)\varphi_b(u)=\delta_{ab} \ .
\ee
Then, the bulk-to-bulk propagator can be written as the usual expansion in the basis of solutions to the boundary problem
\be
 \Sigma(p,u,u_1)=\frac{\Lambda}{(L\Lambda)^{d+1}}\sum_{a=1}^\infty \frac{1}{\mu^2-\mu_a^2} \varphi_a(u)\varphi_a(u_1) \ .
\ee

\subsection{Residues and scattering amplitude}

When the momenta are taken on-shell to one of the masses of the scalar glueballs/mesons $\mu\to \mu_a$, the poles in the bulk-to-bulk propagator are manifest. In the bulk-to-boundary propagator they are implicit in the coefficient $c_\mu$,
\be
 c_\mu\simeq \frac{Y_\nu(\mu_a)}{J_{\nu+1}(\mu_a)(\mu_a-\mu)}=\frac{Y_\nu(\mu_a)}{\sqrt{2}(\mu_a-\mu)}c_a\simeq \frac{\sqrt{2}\mu_aY_\nu(\mu_a)}{\mu_a^2-\mu^2}c_a \ .
\ee
Then, we can approximate the bulk-to-boundary propagator by the single pole contribution
\be
 K(p,u)\underset{p^2\to -M_a^2}{\simeq} \frac{\pi \mu_a^{\nu+1} Y_\nu(\mu_a)}{2^{\nu-\frac{1}{2}}\Gamma(\nu)\Lambda^{\frac{d}{2}-\nu}}   \frac{1}{\mu_a^2-\mu^2} \varphi_a(u) \ ,
\ee
and similarly for the bulk-to-bulk propagator
\be
 \Sigma(p,u,u_1)\underset{p^2\to -M_a^2}{\simeq} \frac{\Lambda}{(L\Lambda)^{d+1}} \frac{1}{\mu^2-\mu_a^2} \varphi_a(u)\varphi_a(u_1) \ .
\ee
Expanding now the normalizable solutions for $u\to 0$, 
\be
 \varphi_a(u)\simeq c_a \frac{\mu_a^\nu}{2^\nu \Gamma(\nu+1)} u^{\frac{d}{2}+\nu}+\ldots \ ,
\ee
and taking into account that $M^2=-p^2$, we obtain 
\bea
 K_\nu(p) &\underset{p^2\to -M_a^2}{\simeq}  & \Lambda^{2(\nu+1)} \frac{\pi  \mu_a^{2\nu+1} Y_\nu(\mu_a)}{2^{2\nu-\frac{1}{2}}\nu \Gamma(\nu)^2}   \frac{c_a}{p^2+M_a^2} \\
 \Sigma_\nu(p,u_1) &\underset{p^2\to -M_a^2}{\simeq}  & -\Lambda^{2+\nu-\frac{d}{2}} \frac{\mu_a^\nu}{2^\nu\nu \Gamma(\nu)L^{d+1}} \frac{c_a}{p^2+M_a^2} \varphi_a(u_1) \ .
\eea
Comparing with the formula \eqref{eq:ZM} we can extract the value of residues in the two-point function of the dual operator, for a glueball/meson of mass $M_a$
\be
 Z_a=\cN \Lambda^{2(\nu+1)} z_a\ , \ z_a= \frac{\pi  \mu_a^{2\nu+1} Y_\nu(\mu_a)}{2^{2\nu-\frac{3}{2}} \Gamma(\nu)^2}c_a \ .
\ee
It is convenient to use this expression for the residue to simplify the formula of the bulk-to-boundary propagator:
\be
 K(p,u)\underset{p^2\to -M_a^2}{\simeq} \Lambda^{\nu+2-\frac{d}{2}}  \frac{2^{\nu-1} \Gamma(\nu) z_a}{ c_a \mu_a^\nu }   \frac{1}{p^2+M_a^2} \varphi_a(u) \ ,
\ee
Combining \eqref{eq:ZM}  and \eqref{eq:K3resHW} we get for the residue of the four-point function
\be
 \cZ_{a; bcd}=-\cN \Lambda^{8-d+4\nu}4^{\nu-1}\Gamma(\nu)^2 v_4 z_b z_c z_d \frac{c_a \mu_a^\nu}{c_b \mu_b^\nu c_c \mu_c^\nu c_d\mu_d^\nu}  \int_0^1 \frac{du_1}{u_1^{d+1}}\varphi_a(u_1) \varphi_b(u_1) \varphi_c(u_1) \varphi_d(u_1) \ .
\ee
Then, the contribution of the quartic term in the scalar potential to the scattering amplitude \eqref{eq:Mabcd} is
\be\label{eq:Mhw}
 \cM_{v_4\, abcd}=-\frac{v_4}{\cN} 4^{\nu-2} \Gamma(\nu)^2\Lambda^{4-d}\kappa_{abcd}\sum_{k=a,b,c,d} \left(\frac{z_{n_1} z_{n_2} z_{n_3}}{z_k}\right)^{1/2} \frac{c_k \mu_k^\nu}{c_{n_1} \mu_{n_1}^\nu c_{n_2} \mu_{n_2}^\nu c_{n_3}\mu_{n_3}^\nu}\Big|_{n_i=a,b,c,d; n_i\neq k} \ ,
\ee
where we have defined the overlap between the solutions of the scalar field as
\be\label{eq:kappa}
 \kappa_{abcd}= \int_0^1 \frac{dv}{v^{d+1}}\varphi_a(v) \varphi_b(v) \varphi_c(v) \varphi_d(v) \ .
\ee
When $v\to 0$ the integrand has the following behavior
\be
\frac{1}{v^{d+1}}\varphi_a(v) \varphi_b(v) \varphi_c(v) \varphi_d(v)\sim v^{d+4\nu-1} \ .
\ee
The integral is convergent as long as $d+4\nu>0$. Unitarity restricts $\nu>-1$, but the condition that the amplitude is finite imposes a stronger restriction for $d< 4$,
\be\label{eq:const}
\nu>-\frac{d}{4}\ \ \Rightarrow \ \ \Delta > \frac{d}{4} \ .
\ee

If instead of a quartic term in the potential of the scalar field we had a term $\sim v_K\Phi^K$, $K>4$, then there would be a contribution to the scattering amplitude similar to \eqref{eq:kappa}, but involving a convolution of $K$ of the modes
\be\label{eq:conv}
 \cM_{v_K}\sim \int_0^1 \frac{dv}{v^{d+1}} \prod_{i=1}^K \varphi_{a_i}(v) \equiv \kappa_K\ .
\ee
This can also be interpreted as the (holographic) wavefunction overlap of all the modes involved in the scattering. This integral is convergent for $(d/2+\nu)K>d$, which gives the condition for a finite $K$-particle scattering amplitude
\be\label{eq:constraint}
 \nu> \frac{d}{K}-\frac{d}{2}\ \ \Rightarrow \ \ \Delta > \frac{d}{K} \ .
\ee
\begin{figure}[H]
\begin{center}
\includegraphics[scale=0.45]{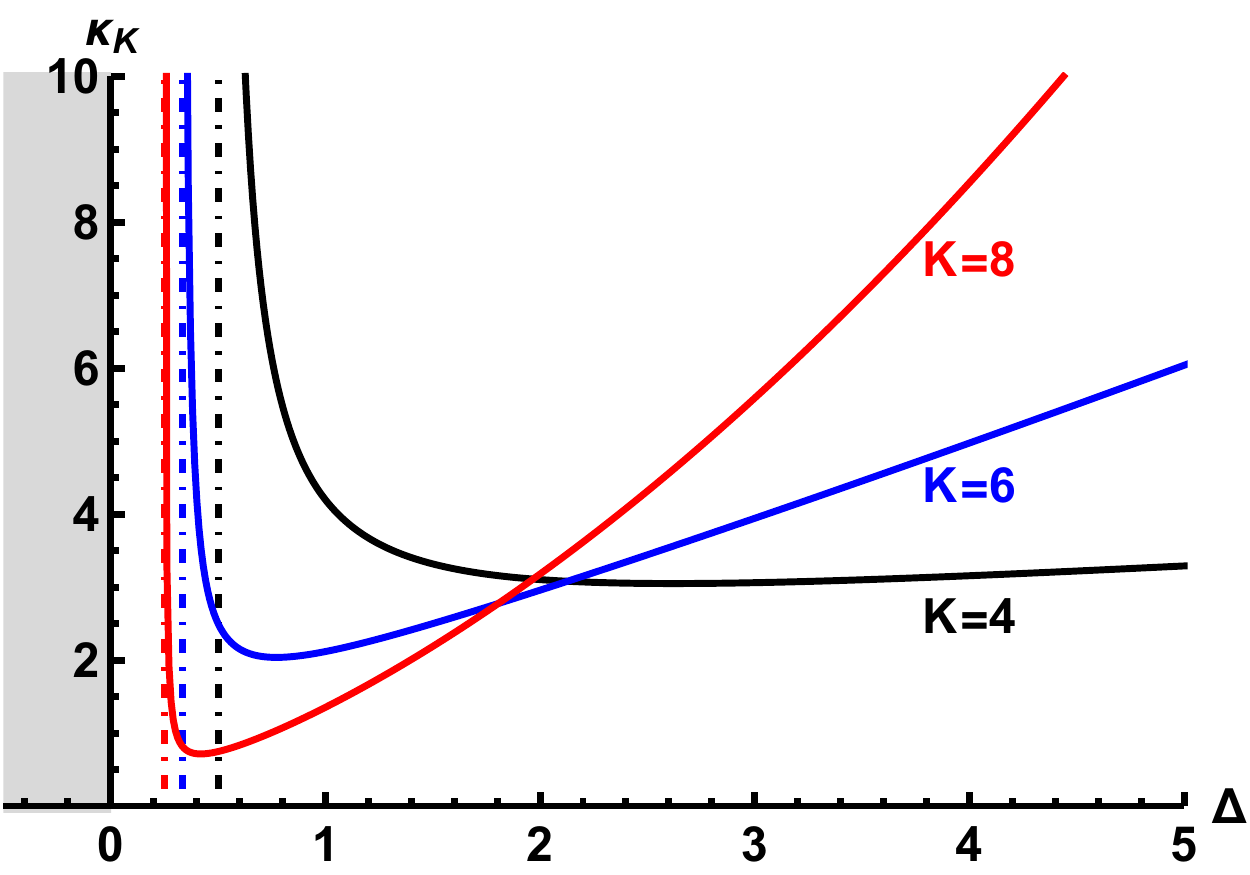}
\includegraphics[scale=0.45]{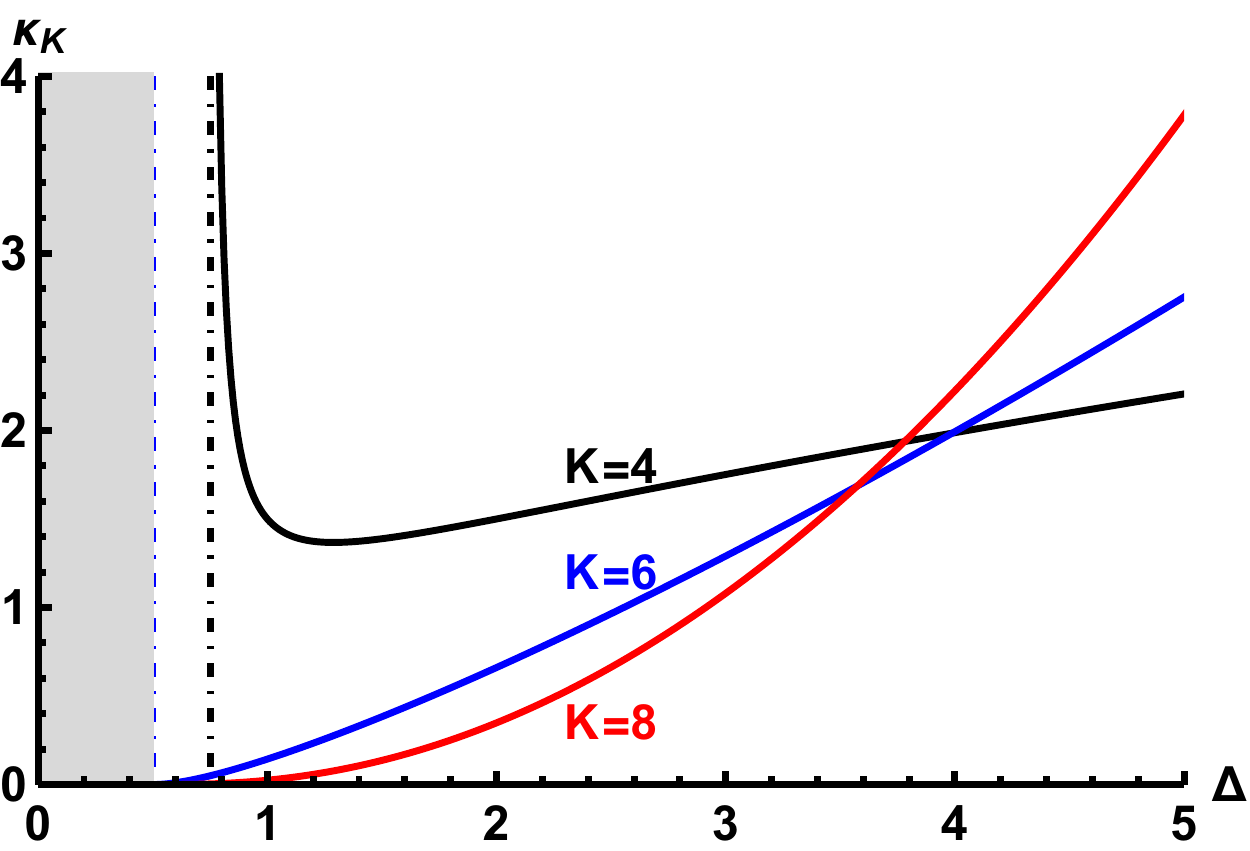}
\includegraphics[scale=0.45]{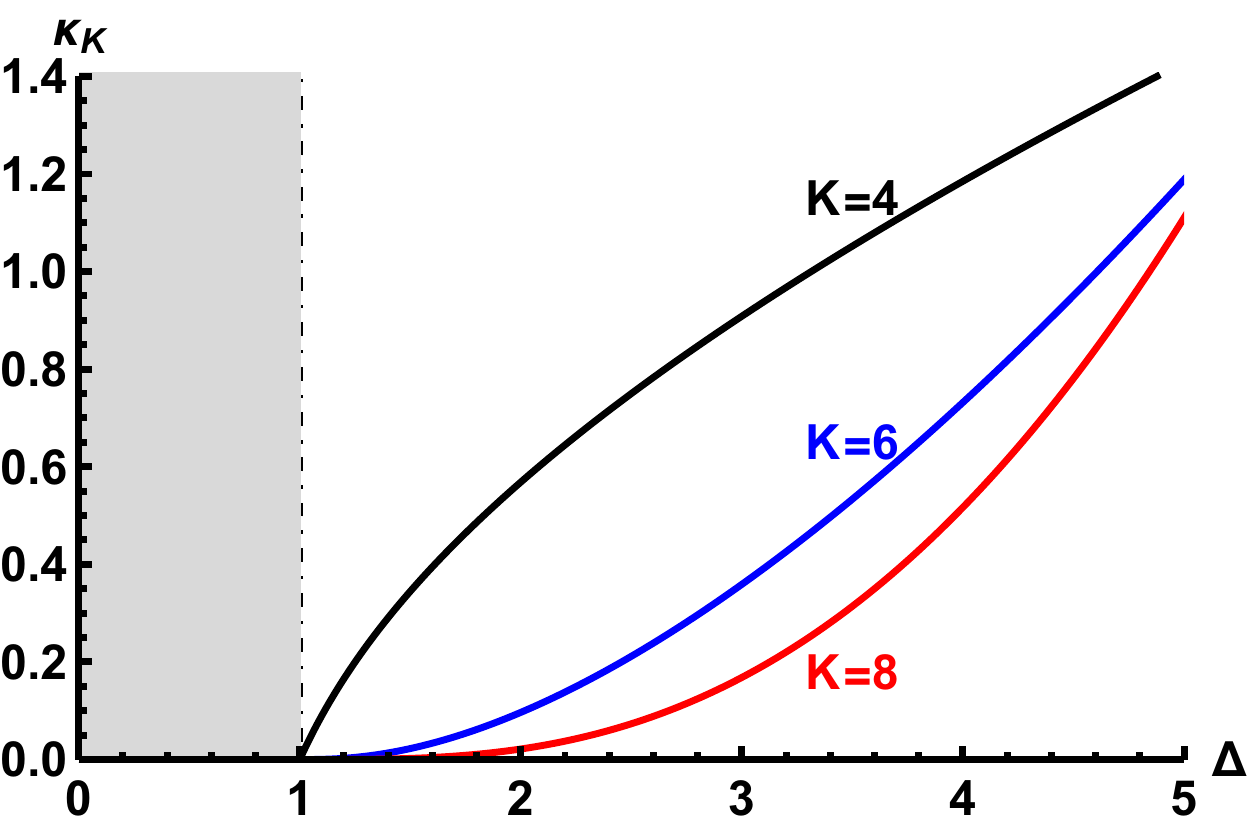}
\caption{We depict the overlap $\kappa_K$ as defined in (\ref{eq:conv}) for various dimensions: $d=2$ (Left), $d=3$ (Middle), and $d=4$ (Right). The three curves correspond to the overlap of $K=4,6,8$ modes. The shaded gray is the region excluded by unitarity (\ref{eq:const}) and the vertical dot-dashed lines are the more stringent constraints (\ref{eq:constraint}) where applicable.}\label{fig:conv}
\end{center}
\end{figure}
We see that larger values of $K$ impose a looser constraint, they become more restrictive than the unitarity bound for $d<2K/(K-2)$. In Fig.~\ref{fig:conv} we have plotted the overlap for various dimensions and for varying number of modes, showing that the corresponding amplitude diverges as the bound (\ref{eq:constraint}) is approached, if the bound is above the unitarity bound. 

\subsection{Scattering length}

We have derived a formula for the scattering length in \eqref{eq:ash} when $d=4$. Since the scattering amplitude \eqref{eq:Mhw} does not depend on the spatial momentum, we can read directly the result for the contribution of the quartic term in the scalar potential in units of the IR scale
\be
 \Lambda a_s(v_4)=\frac{v_4}{16\pi \cN} 4^{\nu-1}\Gamma(\nu)^2 \frac{\kappa_{aabb}}{\mu_a+\mu_b}\left( \frac{|z_a|}{(c_a \mu_a^\nu)^2}+\frac{|z_b|}{(c_b \mu_b^\nu)^2}\right) \ ,
\ee
where
\be\label{eq:overlap}
 \kappa_{aabb}=\int_0^1 \frac{du}{u^5}\varphi_a(u)^2\varphi_b(u)^2 \ .
\ee
This can be further simplified to
\be
 \Lambda a_s(v_4)=\frac{v_4}{32\cN}  \kappa_{aabb}\frac{ \mu_a|Y_\nu(\mu_a)J_{\nu+1}(\mu_a)|+\mu_b|Y_\nu(\mu_b)J_{\nu+1}(\mu_b)|}{\mu_a+\mu_b} \ .
\ee
\begin{figure}[bht!]
\begin{center}
\includegraphics[scale=0.7]{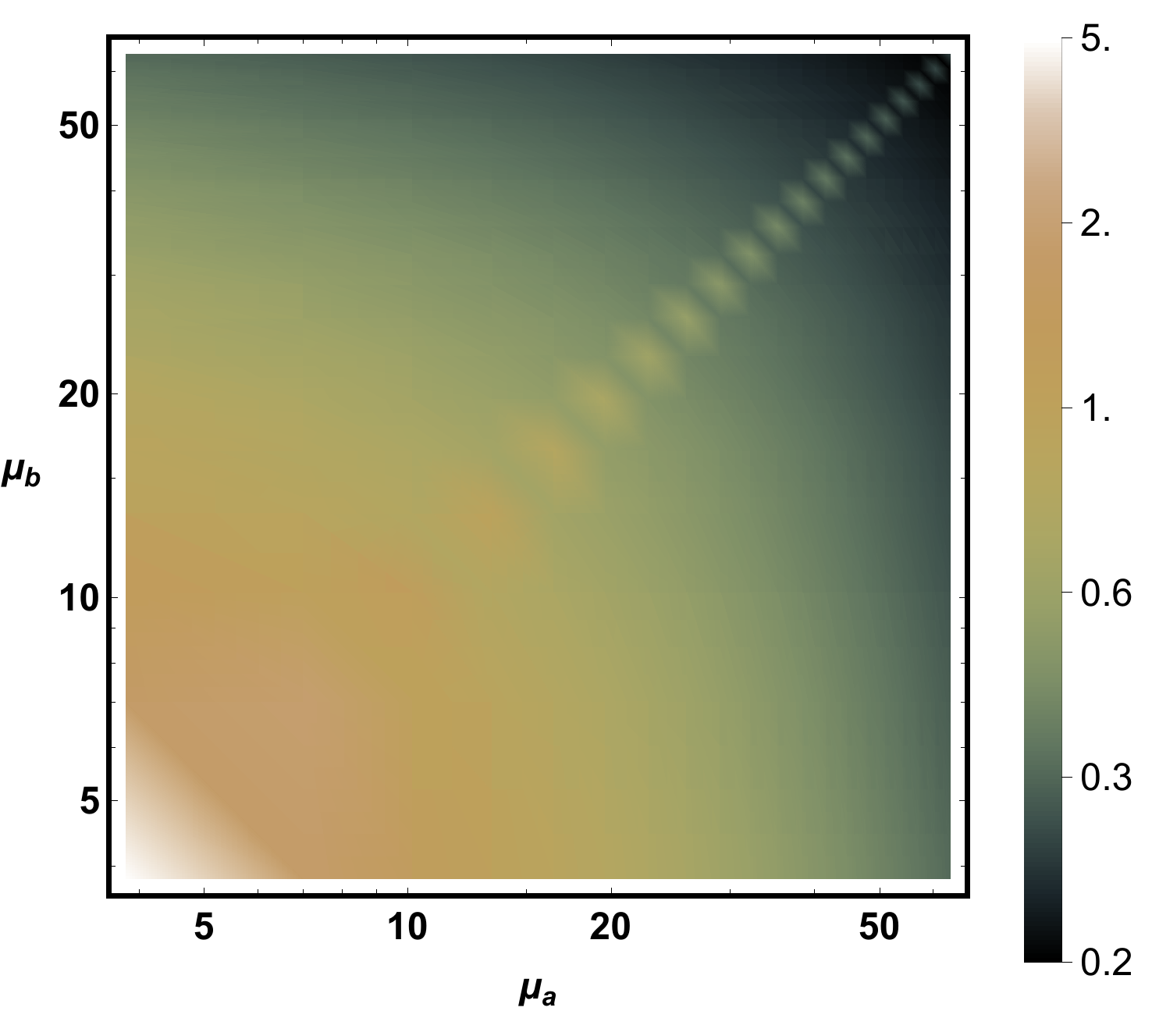}
\caption{Contribution of the quartic term in the scalar potential to the scattering length $10^3\times\cN \Lambda a_s(v_4)/v_4$ as a function of the masses $\mu_a=M_a/\Lambda$, $\mu_b=M_b/\Lambda$ of the particles involved in the scattering for $\nu=1$. Notice that the axes scale logarithmically.}\label{fig:as}
\end{center}
\end{figure}

\begin{figure}[thb!]
\begin{center}
\includegraphics[scale=0.8]{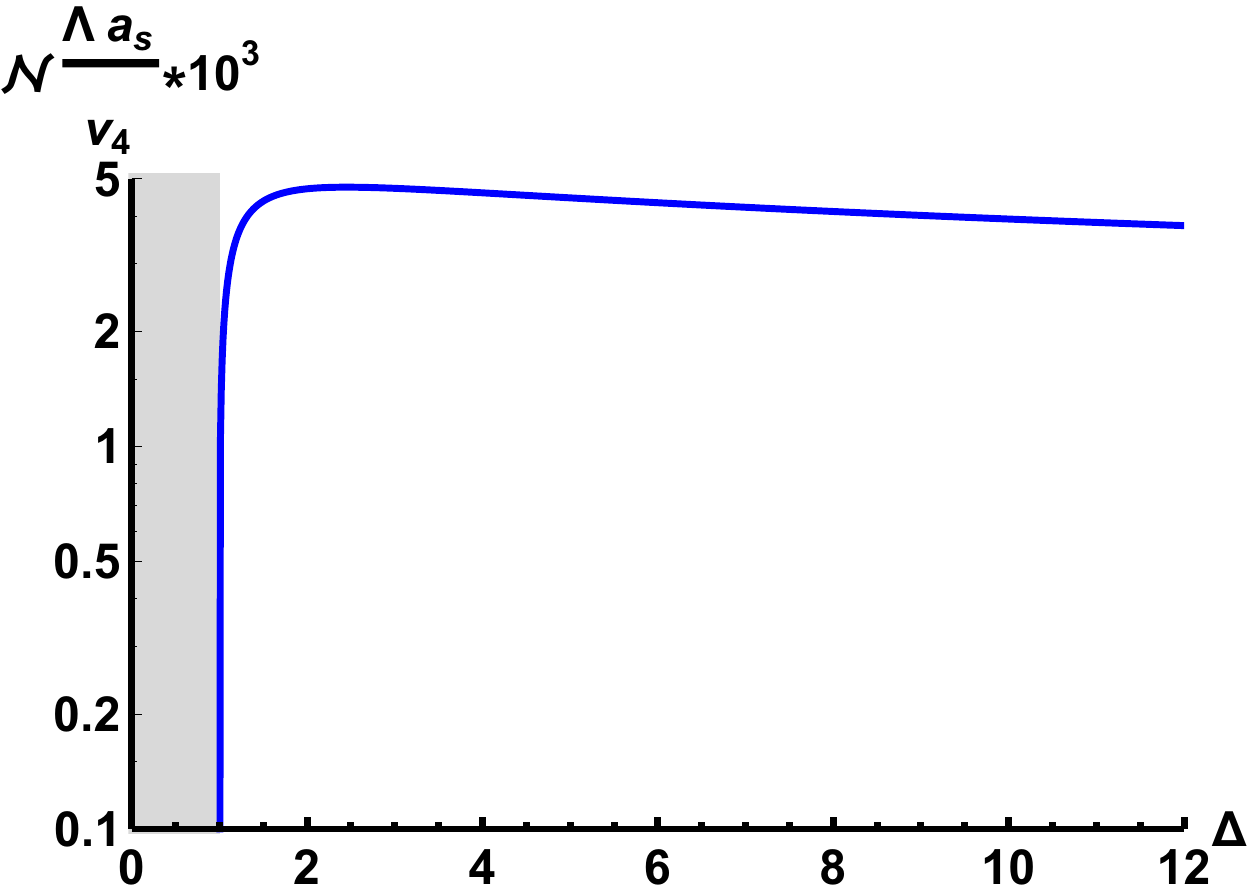}
\caption{We depict the contribution of the quartic term in the scalar potential to the scattering length $10^3\times\cN \Lambda a_s(v_4)/v_4$ for the lowest mass states as a function of the scaling dimension $\Delta$. The shaded gray region is excluded by the unitarity constraint (\ref{eq:const}), which in this case is no weaker than (\ref{eq:constraint}).}\label{fig:as2}
\end{center}
\end{figure}

The overlap (\ref{eq:overlap}) is clearly positive $\kappa_{aabb}>0$, so the sign of the contribution to the scattering length is determined by the sign of the coefficient $v_4$ in the quartic term of the potential. For a large enough value of $|v_4|$ this contribution would dominate and determine the sign of the full scattering length. In this case the interaction between the scalar glueballs or hadrons associated to the operator $\cO$ would be repulsive for $v_4>0$ and attractive for $v_4<0$. In general we expect the overlap to decrease when the mass difference is larger, as the solutions for the scalar field associated to each mass will be peaked at different positions in the radial direction. A smaller scattering length implies weaker interactions between states of different mass. We have confirmed the expected trend of $a_s(v_4)$ by computing numerically the scattering length for a large set of masses; see Fig.~\ref{fig:as}. 
In Fig.~\ref{fig:as2} we, on the other hand, show how the scattering length $a_s$ depends on the scaling dimension of the scalars involved in the process. For small scaling dimension the scattering length increases, reaching a maximum and then slowly decreasing indefinitely.

\section{Discussion}\label{sec:discuss}

We have presented a method to compute the scattering length of the effective interaction between color singlet states (glueballs or hadrons) in a large-$N$ strongly coupled gauge theory with a holographic dual. We have focused on the states created by a scalar operator of fixed dimension $\Delta$ and performed explicitly the calculation in a hard wall model, in the spirit of holographic QCD. It is, in principle, straightforward to extend our analysis to operators of different spins or to consider the scattering process between particles created by operators of different dimensions.

In the hard wall example we have found that the condition of having a finite amplitude puts a lower bound on the dimensions of the scalar operator. The constraint stems from the properties of the solution in the asymptotic region, so we expect it to be universal for any model that is asymptotically $AdS$. We have argued that the bound is above the unitarity bound for theories in $d< 2K/(K-2)$ spacetime dimensions if there is a polynomial term of order $K$ in the scalar potential of the dual theory. Note that for the case we have worked out explicitly, $K=4$, the bound is above the value of scalar operators in some CFTs. For instance, in the $d=3$ Ising model $\Delta_\sigma \simeq 0.518$ (see, {\emph{e.g.}}, \cite{Pelissetto:2000ek}), which is below $3/4$ bound that we found in the holographic model. The same holds for the $d=2$ Ising model, where $\Delta_\sigma=1/8$, while the holographic bound is $1/4$. 

It is interesting to ask why the bound we are proposing is stronger than expected? It might be that the bound only applies to a restricted set of theories with holographic duals, in the large-$N$ limit, or with a discrete spectrum of massive states. It would be especially interesting if the last option was true, as it would imply that a theory with an operator of low enough dimension cannot have a discrete gapped spectrum, going beyond conformal bootstrap bounds (see, {\emph{e.g.}}, \cite{Poland:2018epd} for a review). Note that theories with a holographic dual that do not have a fixed point in the UV, {\emph{i.e.}} with an asymptotic geometry different from $AdS$, might also avoid the bound. However, this will signal UV physics different from that of an ordinary field theory in $d$ dimensions.

\begin{figure}[thb!]
\begin{center}
\includegraphics[scale=0.3]{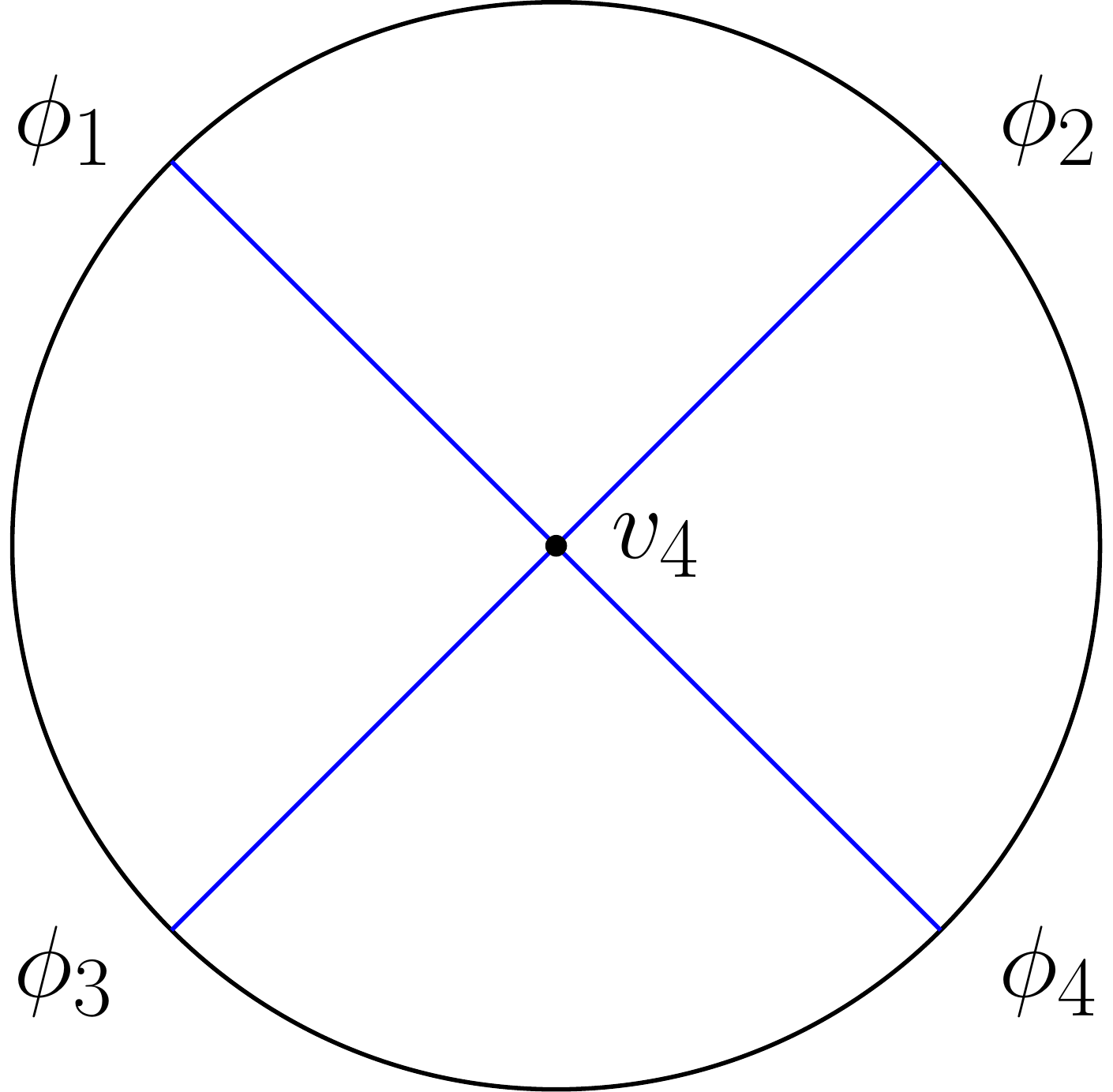}
\qquad\qquad
\includegraphics[scale=0.3]{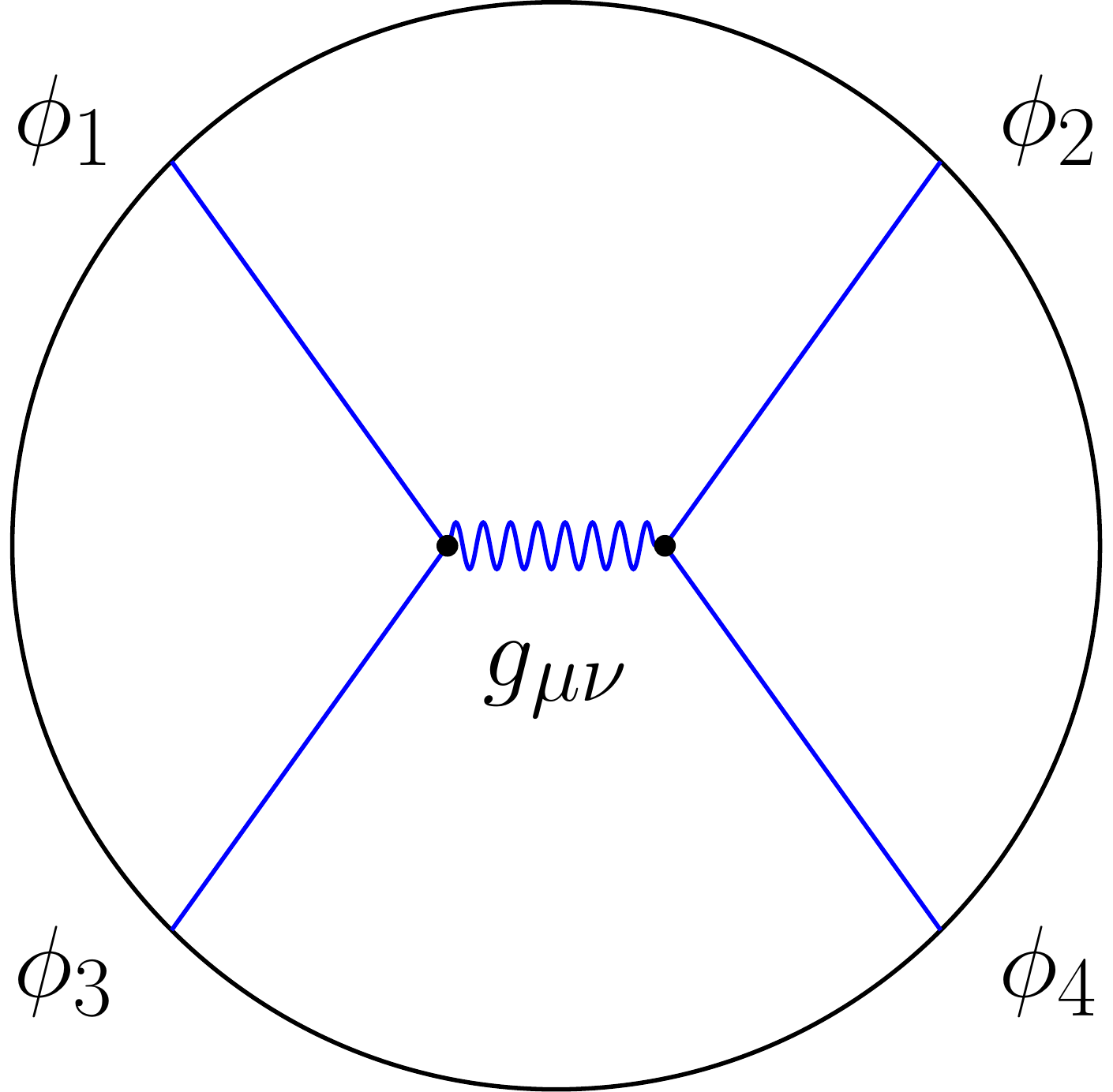}
\caption{Witten diagrams corresponding to the contribution by the quartic interaction that we have computed (Left) and the contribution due to one graviton exchange (Right).}\label{fig:wittendiag}
\end{center}
\end{figure}

As our purpose was to point out the possibility of computing this observable in holography and to expose the method, we have restricted for simplicity to interactions that only involve the dual scalar field. Our calculation can be understood pictorially as a Witten diagram with four scalar legs joining at a single point in the bulk. We should mention that the calculation of four-point functions of scalar operators through Witten diagrams and other techniques in AdS/CFT has a long history, starting with \cite{Muck:1998rr,Freedman:1998bj}. However, previous works have largely focused on conformal theories and using a representation in position space, or Mellin transforms that take advantage of conformal symmetry \cite{Mack:2009gy,Penedones:2010ue,Fitzpatrick:2011ia,Paulos:2011ie}. In some cases, but still restricted to conformal theories, correlators in momentum space have also been studied; for recent works, see \cite{Albayrak:2018tam,Albayrak:2019yve,Bzowski:2019kwd}.

It should be noted that the scalar field is coupled to the metric and maybe other fields as well, so there are more contributions to the effective interaction of scalar states and therefore to the scattering length. In particular, there should be a universal contribution corresponding to the exchange of one graviton between the two scalars, as in Fig.~\ref{fig:wittendiag}. It seems quite likely that the graviton exchange will produce a tensor contribution to the effective potential, analogous to the one observed in the nucleon-nucleon potential. It would be quite interesting to derive the effective potential for fermion fields and compare with fits of the nucleon-nucleon potential to experiments. Exchange diagrams would also introduce additional momentum dependence in the scattering amplitude, whereas the contact interaction in the effective potential we have studied in this paper reduces to a delta function in space.

Following this last line of thought, a direct comparison with QCD would be more consequential for holographic models that are designed to approximate it as closely as possible, such as the IHQCD or V-QCD models \cite{Gursoy:2010fj,Jarvinen:2011qe}. It should be noted that baryons usually do not have a simple description in the holographic model as fields in the gravity action, although it is possible to avoid this issue by considering different types of large-$N$ limits \cite{Corrigan:1979xf,Armoni:2003gp,HoyosBadajoz:2009hb,Hoyos:2016ahj}. 

Another interesting extension of our work would be to find the scattering amplitude for states in a theory with a known string theory dual. Our analysis requires that the theory is ``confining'', in the sense that the spectrum should be discrete; a few examples were mentioned in the introduction. In these type of theories, the gravity dual has a geometry that either ends in a singular manner or smoothly if a cycle in the compact space collapses to zero size. Both of these cases are quite different from the Dirichlet condition imposed in the hard wall model. Presumably the difference will lead to drastic ramifications of the IR physics, and so will affect the scattering length. It would thus be important to compute the scattering length in these other bulk realizations of gapped spectra and compare them to the case at hand, and to discern any emerging qualitative differences.

\paragraph{Acknowledgements}

We would like to thank Kari Rummukainen for discussions. C.~H. is partially supported by the Spanish grant PGC2018-096894-B-100 and by the Principado de Asturias through the grant FC-GRUPIN-IDI/2018/000174. N.~J. is supported in part by the Academy of Finland grant no. 1322307.

\bibliographystyle{JHEP}
\bibliography{refsscat}

\end{document}